\documentclass{aipproc}
\pdfoutput=1
\layoutstyle{8x11single}
\usepackage{xspace}
\usepackage{units}
\usepackage{subfig}
\usepackage{amssymb,amsmath}
\newcommand{\ccpip}{CC1\ensuremath{\pi^+}\xspace}
\newcommand{\ccpi}{CC1\ensuremath{\pi^0}\xspace}
\newcommand{\ncpi}{NC1\ensuremath{\pi^0}\xspace}
\newcommand{\ud}{\ensuremath{\mathrm{d}}}
\newcommand{\diff}[1]{\ensuremath{\ud \sigma / \ud #1}}
\newcommand{\mb}{MiniBooNE\xspace}

\title{Comparing pion production models to \mb data}
\author{P.\ A.\ Rodrigues}{address={Department of Physics and
    Astronomy, University of Rochester, Rochester, NY, USA}}
\begin{document}
\begin{abstract}
  Predictions for neutrino-induced charged- and neutral-current single
  pion production on $\mathrm{CH_2}$ from theoretical models and Monte
  Carlo event generators are compared with the cross section measurements from the
  \mb experiment.
\end{abstract}

\maketitle

Improved understanding of neutrino-induced single pion production in
interactions on nuclei in the $E_\nu \lesssim \unit[1]{GeV}$ region is
important for current and upcoming neutrino oscillation experiments,
as these processes form backgrounds to both $\nu_\mu \to \nu_e$
appearance and $\nu_\mu \to \nu_\mu$ disappearance
signals. Single pion production on nuclei is also interesting in its
own right, as it probes features such as the contribution of
nonresonant background terms; possible multinucleon contributions to
the cross section; and the effect of final state interactions on both
the total cross section and differential cross sections in pion
kinematic variables.

A key component in answering these questions, and reducing systematic
uncertainties for oscillation experiments, is detailed measurement of
pion production cross sections on nuclei in a range of channels. The
cross section for pion production on individual nucleons was measured
in deuterium bubble chambers at ANL~\cite{anl-spp} and
BNL~\cite{bnl-spp} for $E_\nu$ around \unit[1]{GeV}, but these
measurements disagree with one another at the 30--40\% level, and both
have large uncertainties. More recent measurements, taken on plastic
scintillator and water, have been made by the K2K~\cite{k2k-ccpi0,k2k-nc1pi0,k2k-cc1pip}
experiment, although these are not presented as absolute cross
sections.

At NuInt09, a detailed set of comparisons was made between the
predictions of a wide range of models for charged current single pion
production on carbon~\cite{nuint09-comparisons}. The present study
considers a smaller number of models, but includes comparisons to \mb
data which were not available at the time of NuInt09.

\section{\mb data}
The most detailed measurements to date of single pion production on
nuclei have been made by the \mb experiment, which produced
high-statistics absolute cross sections for charged current single
$\pi^+$ (\ccpip)~\cite{mb-cc1pip}, charged current single $\pi^0$
(\ccpi)~\cite{mb-cc1pi0}, and neutral current single $\pi^0$ (\ncpi)
production~\cite{mb-nc1pi0}. For each of these measurements, a range
of differential cross sections in relevant kinematic quantities has
been provided, facilitating thorough comparison with model
predictions. In each case the signal is defined by the particles
exiting the nucleus, rather than particles produced at the interaction
vertex, so the results are able to test the combination of nucleon-level cross
section, nuclear effects, and final state interactions.

The \mb data are taken in the Booster neutrino beam, which consists
mostly of muon neutrinos, and has a peak neutrino energy of around
\unit[600]{MeV}. The neutrino flux is small beyond about
\unit[2]{GeV}~\cite{mb-flux}. Differential cross sections provided by \mb are
averaged over this flux.

The three datasets published by \mb are summarized below:
\begin{description}
\item[\ccpip] The signal is defined as a muon and exactly one $\pi^+$
  exiting the nucleus, with no other mesons, but any number of
  nucleons. Resonant and coherent $\pi^+$ production can contribute to
  this sample.
\item[\ccpi] The signal is defined as a muon and exactly one $\pi^0$
  exiting the nucleus, with no other mesons, but any number of
  nucleons. Resonant $\pi^0$ production contributes to this sample,
  along with a potentially significant fraction of resonant $\pi^+$
  events in which the $\pi^+$ charge exchanges in the nucleus. For
  this sample, \mb present differential distributions averaged over
  the neutrino flux from 0.5 to \unit[2]{GeV}.
\item[\ncpi] The signal is defined as no muon and exactly one $\pi^0$
  exiting the nucleus, with no other mesons, but any number of
  nucleons. Resonant and coherent $\pi^0$ production are expected to
  be the main contributions to this sample. Measurements have also
  been made in the $\bar{\nu}_\mu$ flux, but are not considered here.
\end{description}

\section{Models used}
The models compared in this work can be broadly divided into two
categories, namely theoretical models, and full Monte Carlo (MC) generators. While
the theoretical models produce total or differential cross sections
for a given process as output, generators produce samples of events
suitable for use as input to a full experimental
simulation. Predictions for each of the models were kindly provided by
the relevant groups.

The theoretical models considered are:
\begin{description}
\item[Athar \emph{et al.}] This model uses a $\Delta$-dominance
  assumption with a local density approximation for the nuclear
  effects. In-medium modifications to the $\Delta$ properties are
  included, along with an MC cascade model of final state
  effects~\cite{athar1,athar2}.
\item[GiBUU] The GiBUU code provides a unified treatment of nuclear
  transport processes for interactions of nucleons, nuclei, pions,
  electrons and neutrinos with nuclei. The code provides a careful
  treatment of particle transport through the nucleus via the BUU
  equation. The neutrino-nucleon cross sections for single pion
  production account for 13 resonances with vector couplings taken
  from the MAID model for pion photo- and electroproduction, and axial
  couplings from PCAC. Nonresonant background terms are also included~\cite{gibuu}.
\item[Nieves \emph{et al.}] The model of Nieves \emph{et al.}
  describes both quasielastic and single pion production processes in
  the region around \unit[1]{GeV}. The single pion production
  description is based on the SU(2) nonlinear $\sigma$ model, with
  nonresonant background terms included, and only the $\Delta$
  resonance considered. The parameters $C_5^A(0)$ and
  $M_A^\Delta$ are tuned to ANL and BNL data, while the nuclear
  effects are taken from photon, electron and pion interactions with nuclei~\cite{nieves}.
\end{description}

The Monte Carlo generators considered are:
\begin{description}
\item[GENIE 2.6.2] The Rein-Sehgal model is used for values of the
  hadronic invariant mass $W < \unit[1.7]{GeV}$, with 16 resonances
  included. Nuclear effects are treated using the relativistic Fermi
  Gas model~\cite{genie}.
\item[NEUT 5.1.4.2] NEUT also uses the Rein-Sehgal model, for $W <
  \unit[2]{GeV}$, along with the relativistic Fermi Gas model. The
  full set of 18 resonances included used in the Rein-Sehgal paper is
  included~\cite{neut}.
\item[NuWro] NuWro explicitly considers only the $\Delta$ resonance,
  with heavier resonances treated using quark-hadron duality from the
  DIS cross section. The spectral function is employed for nuclear
  effects~\cite{nuwro}.
\end{description}
All three MC generators use their own cascade model for final state
effects.

At the time of writing, not all of the models had produced predictions
for all of the distributions. Each figure shows all the distributions
that were available.

\begin{ltxfigure}[tp]
  \centering
  \includegraphics[width=0.33\textwidth]{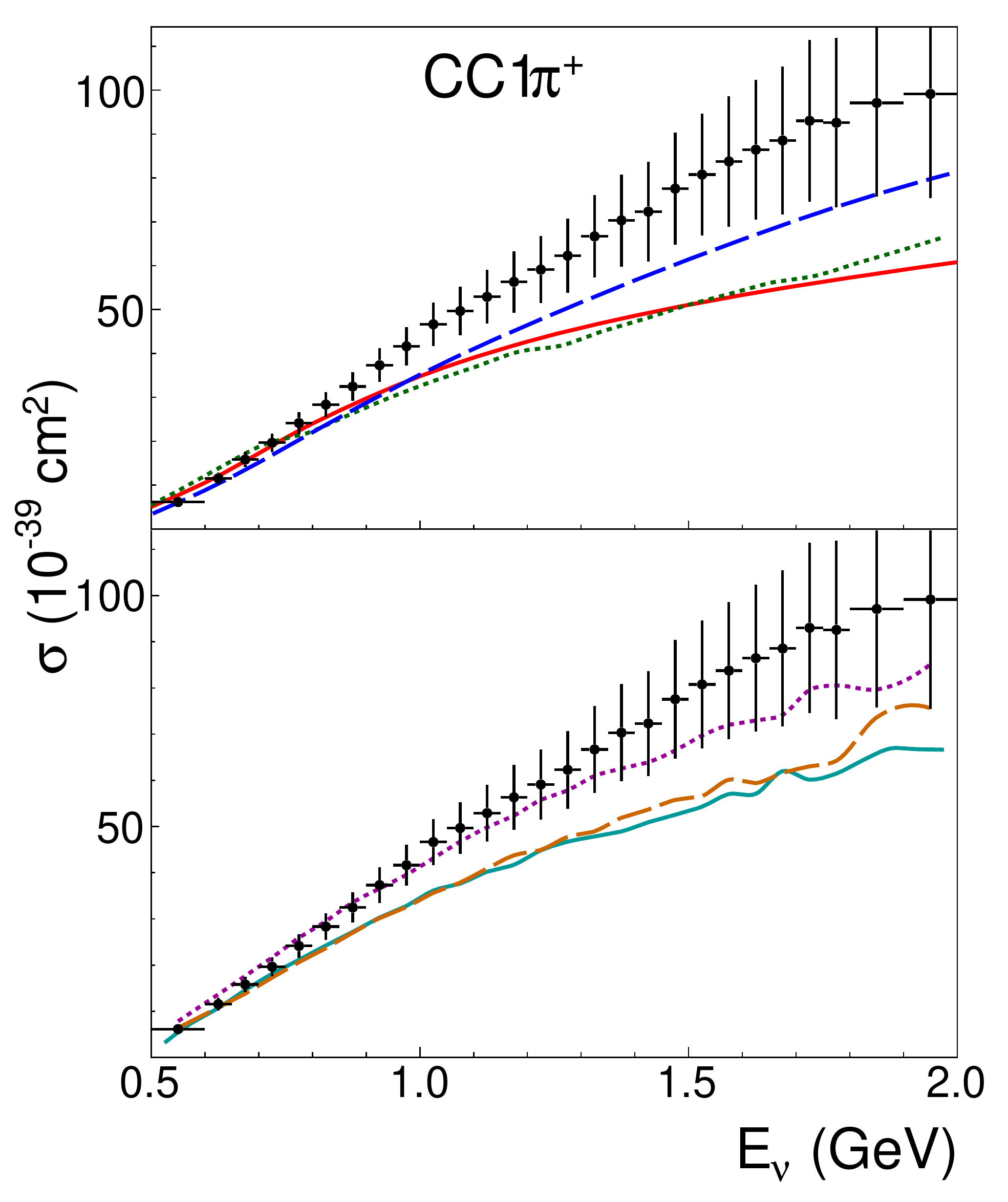}
  \includegraphics[width=0.33\textwidth]{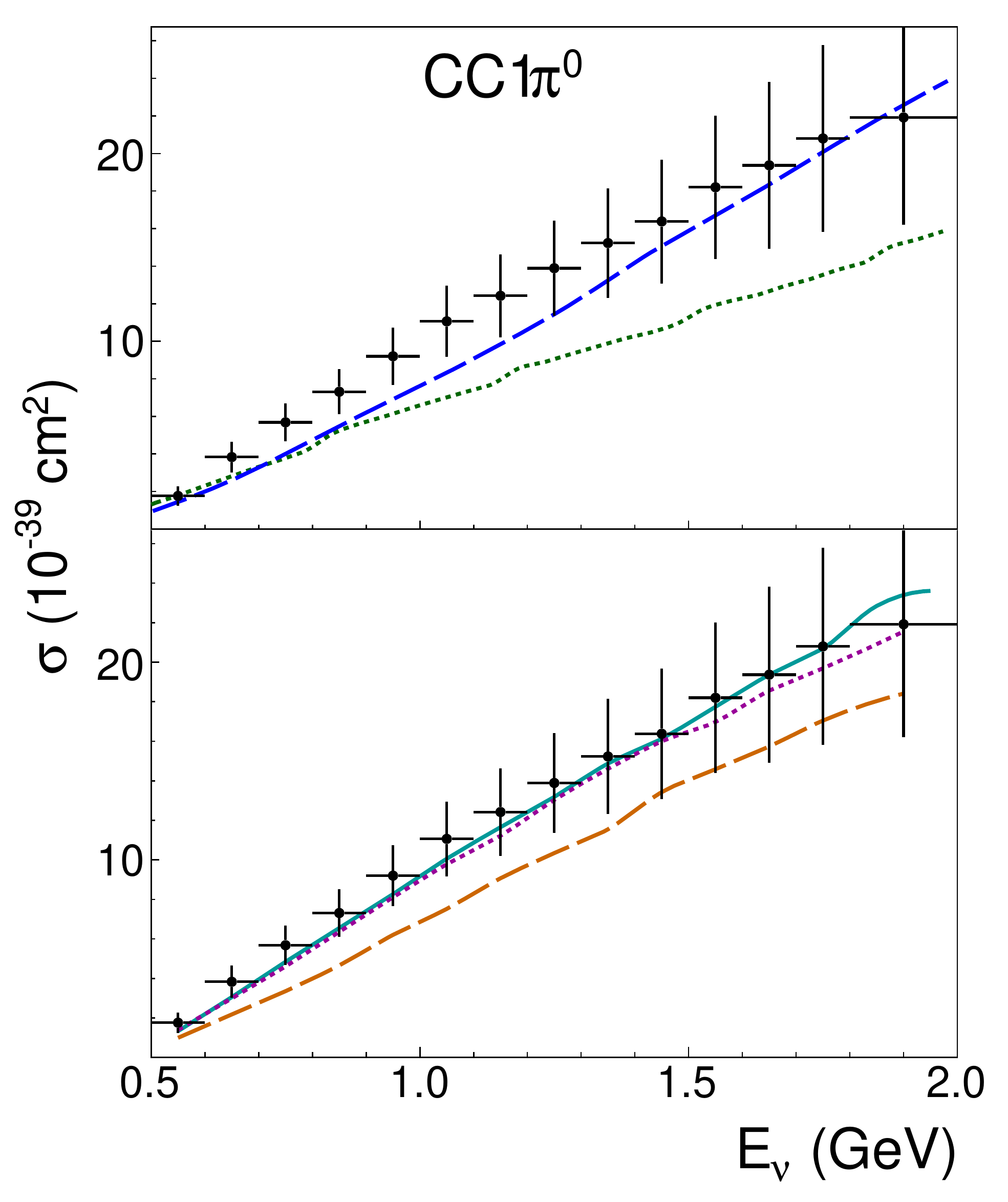}\\
  \includegraphics[width=0.5\textwidth]{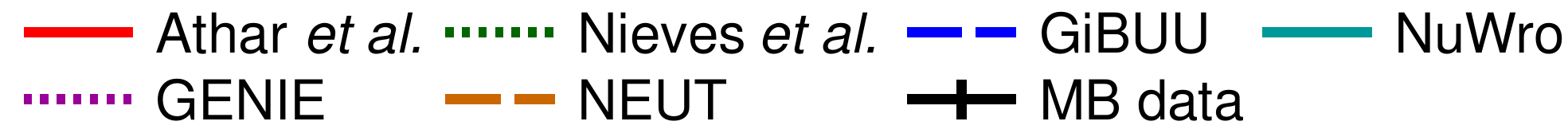}
  \caption{\ccpip (left) and \ccpi (right) total cross sections as a
    function of neutrino energy for each of the models and for the \mb
    data. The top row shows the theoretical model predictions, while
    the bottom row shows the MC generator predictions. (The data is
    the same top and bottom.)}
  \label{fig:totalxsec}
\end{ltxfigure}
 
\begin{ltxfigure}[tp]
  \centering
  \includegraphics[width=0.33\textwidth]{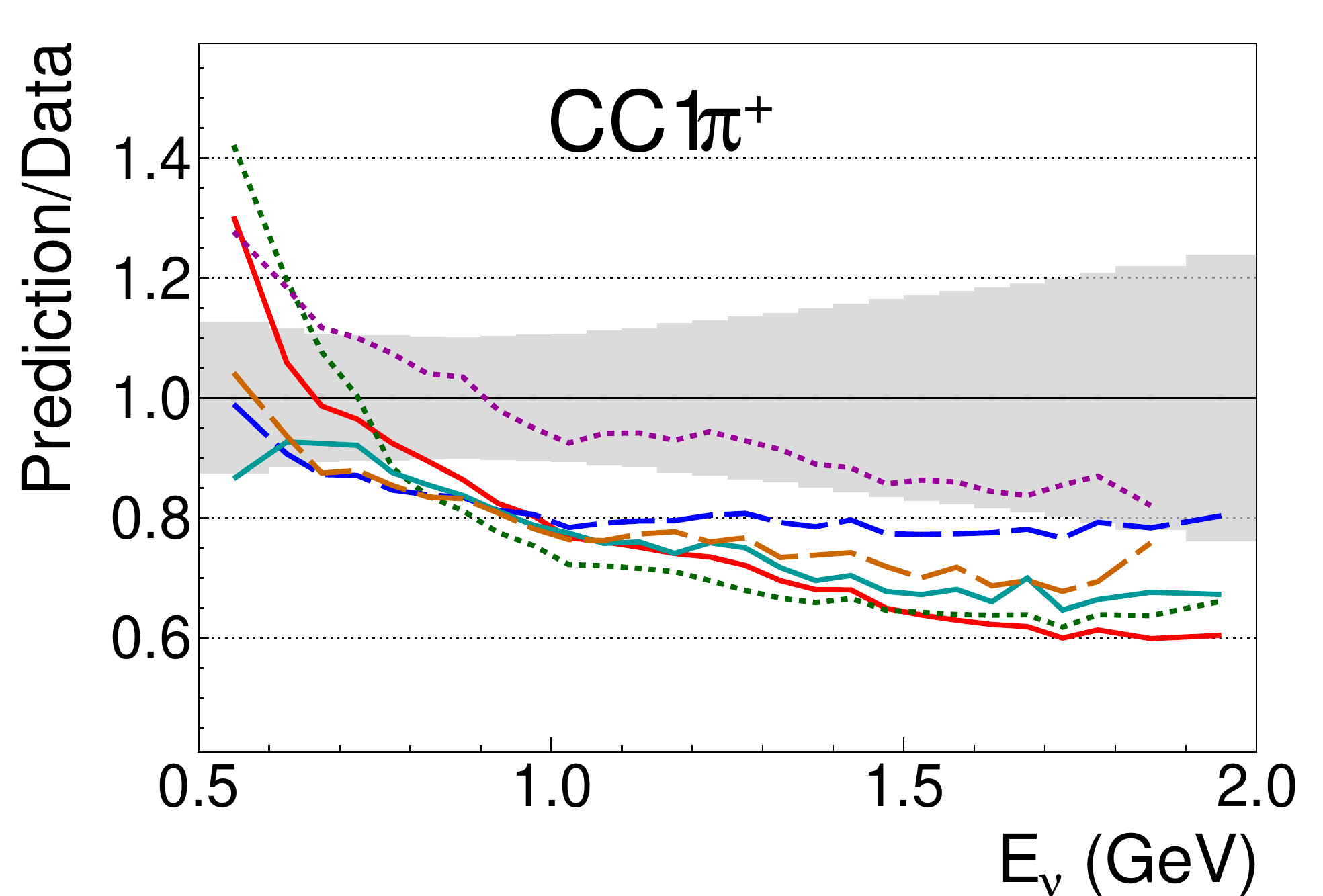}
  \includegraphics[width=0.33\textwidth]{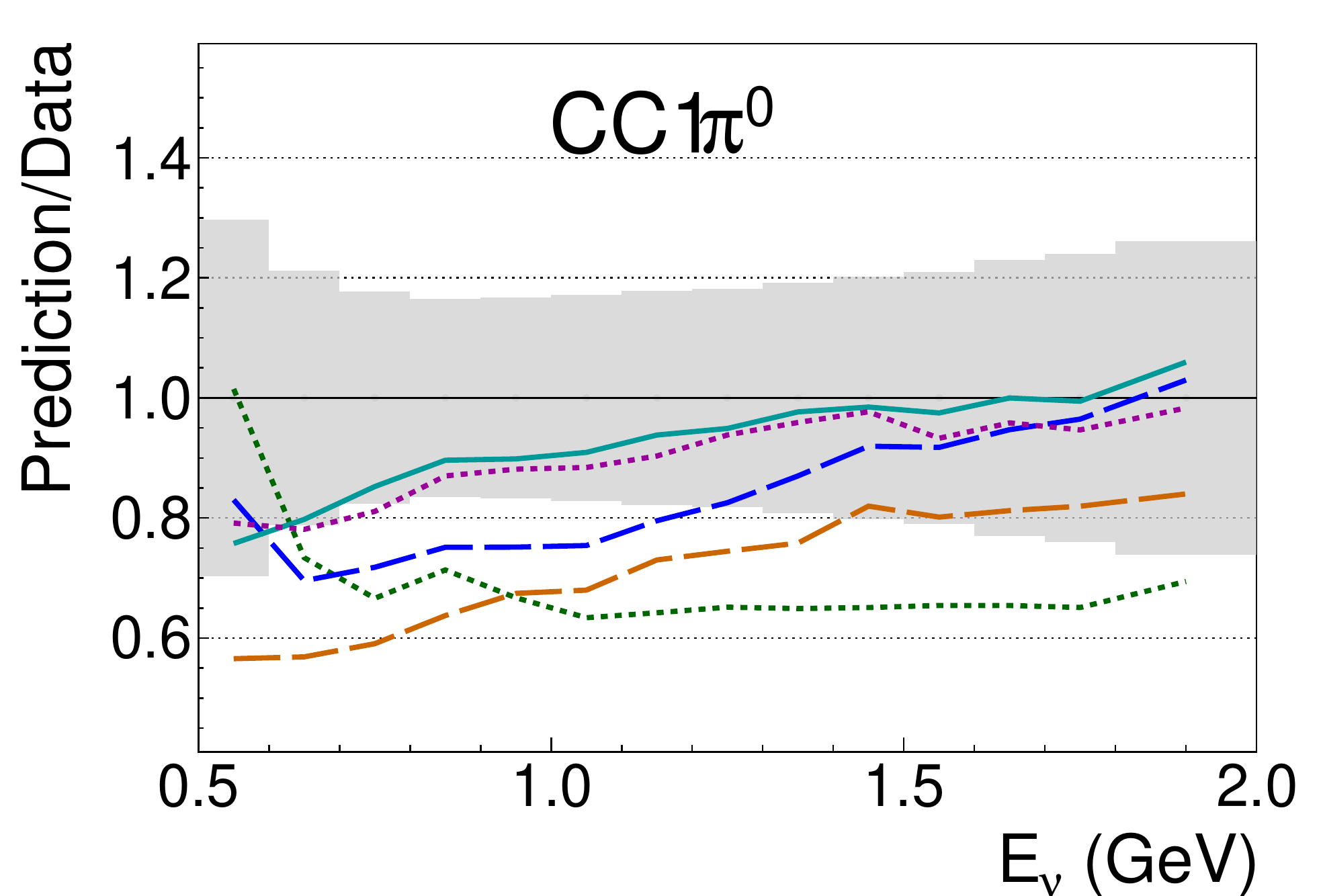}\\
  \includegraphics[width=0.5\textwidth]{horiz_legend}
  \caption{Ratios of prediction to \mb data in the total cross section as
    a function of energy for \ccpip (left) and \ccpi (right). The grey
  shaded region shows the size of the data uncertainty.}
  \label{fig:enu-ratios}
\end{ltxfigure}

\section{Comparisons}

Figure~\ref{fig:totalxsec} shows the total cross sections as a
function of energy for \ccpip and \ccpi. The predictions generally
fall below the data in the case of \ccpip, with GENIE coming
closest. In \ccpi, the agreement with data is, on the whole, better,
although there is still some tendency for predictions to fall below
the data points.

Figure~\ref{fig:enu-ratios} shows the ratio of total cross section
prediction to data as a function of energy, for the two CC samples. A
few features stand out. Firstly, although the predictions differ
significantly in the overall normalization of the cross section, the
shapes of the prediction/data ratios, particularly in \ccpip, are
quite similar, being higher at low energies than at higher
energies. Secondly, although the \ccpi ratios also show similarities
between the predictions, the shapes are quite different from the
\ccpip shapes. This suggests that the discrepancies between the
predictions and the \mb data cannot be solely due to a mismodelling of
the flux shape, which would lead to similar shapes in the two
cases.

The ability to draw conclusions such as these is one of the advantages of
taking the three \mb pion production data sets together. Less
optimistically, the difficulty of obtaining simultaneous agreement
with the two CC samples highlights a potential problem for other
experiments (especially neutrino oscillation experiments) in using the
\mb data to constrain cross section predictions. An example of this is
the T2K oscillation analysis, which introduced \emph{ad hoc} tuning
parameters with large uncertainties to reflect the difficulty of
obtaining simultaneous agreement with the three \mb data
sets~\cite{me-nufact}.

\begin{ltxfigure}
  \centering
  \includegraphics[width=0.33\textwidth]{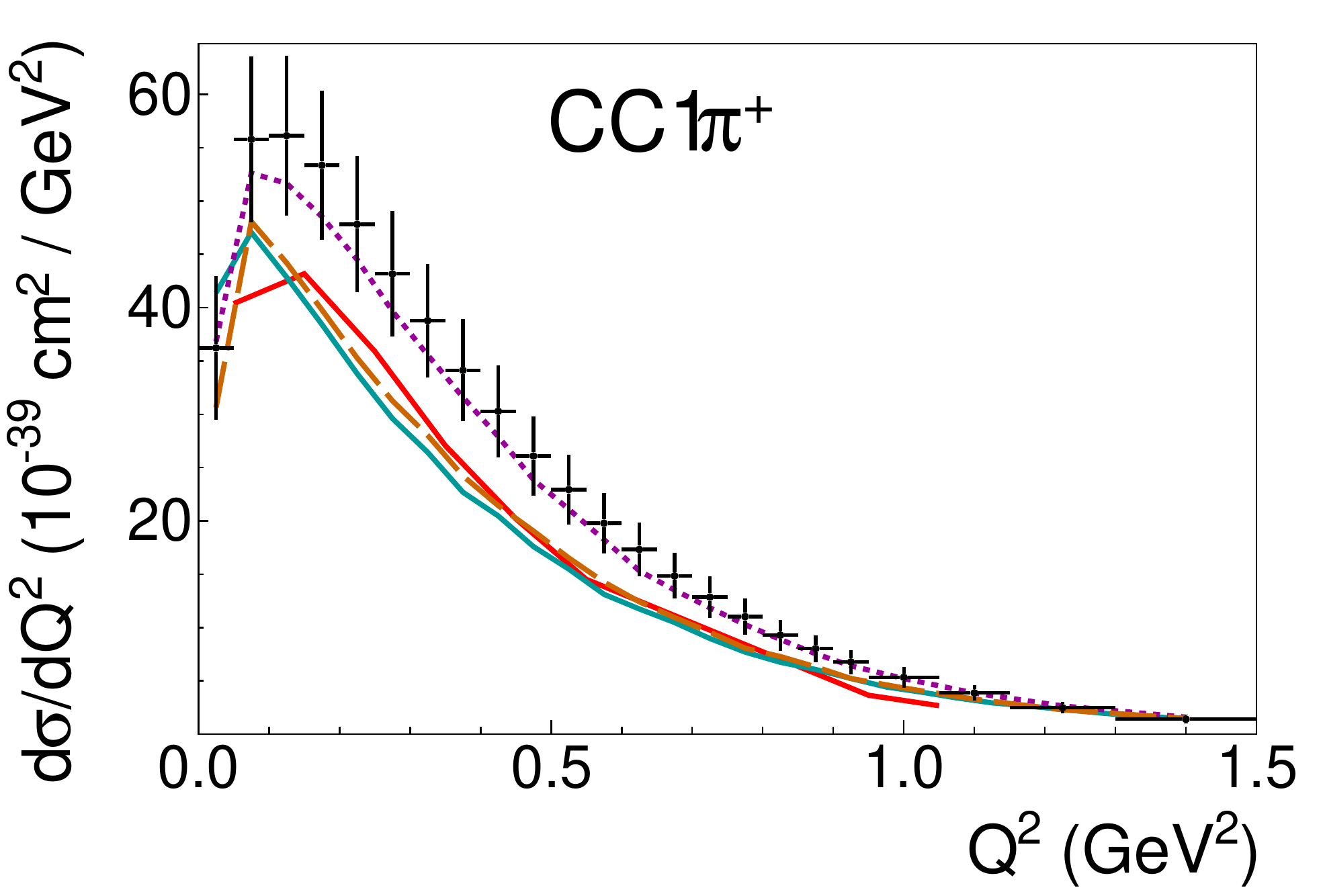}
  \includegraphics[width=0.33\textwidth]{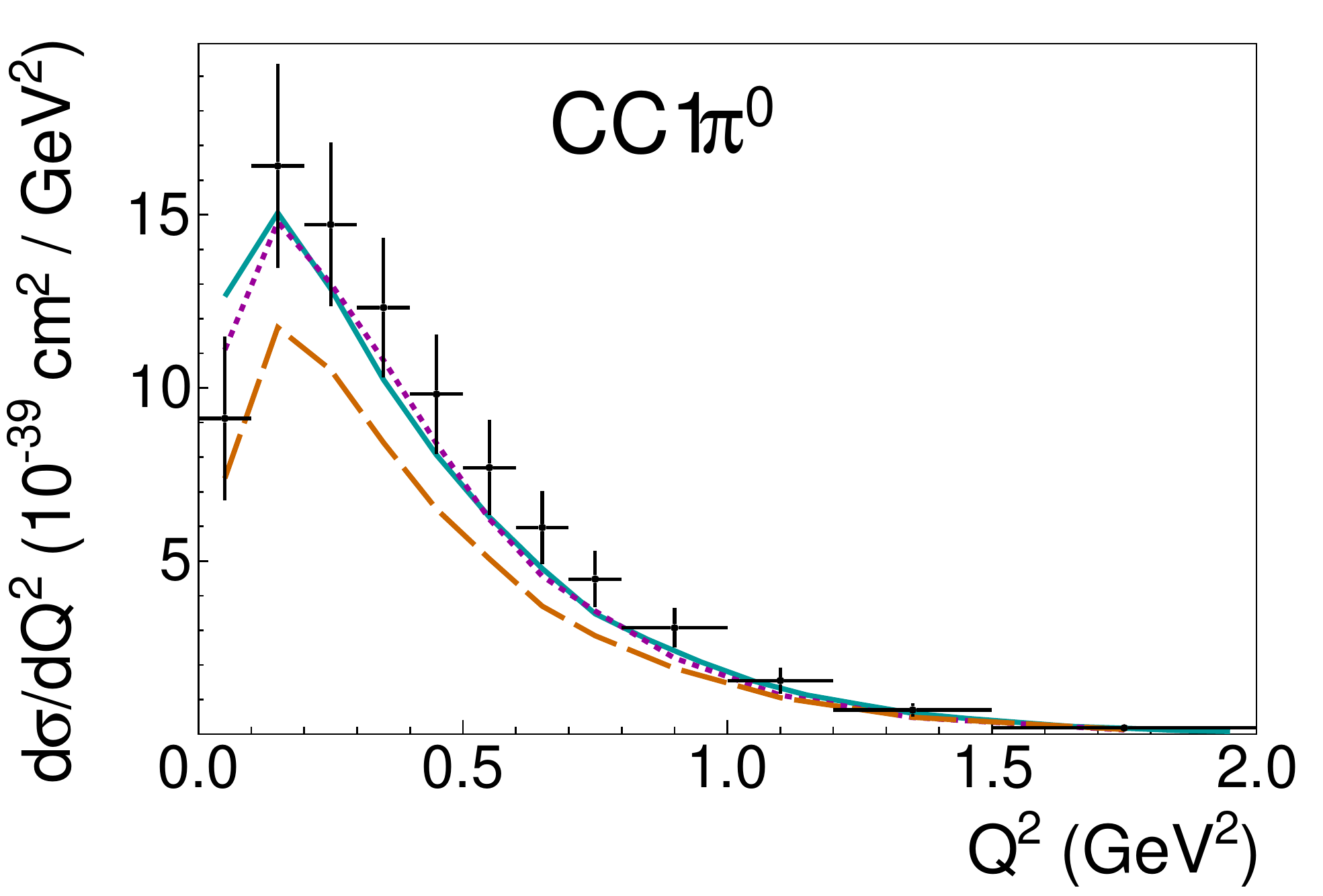}\\
  \includegraphics[width=0.33\textwidth]{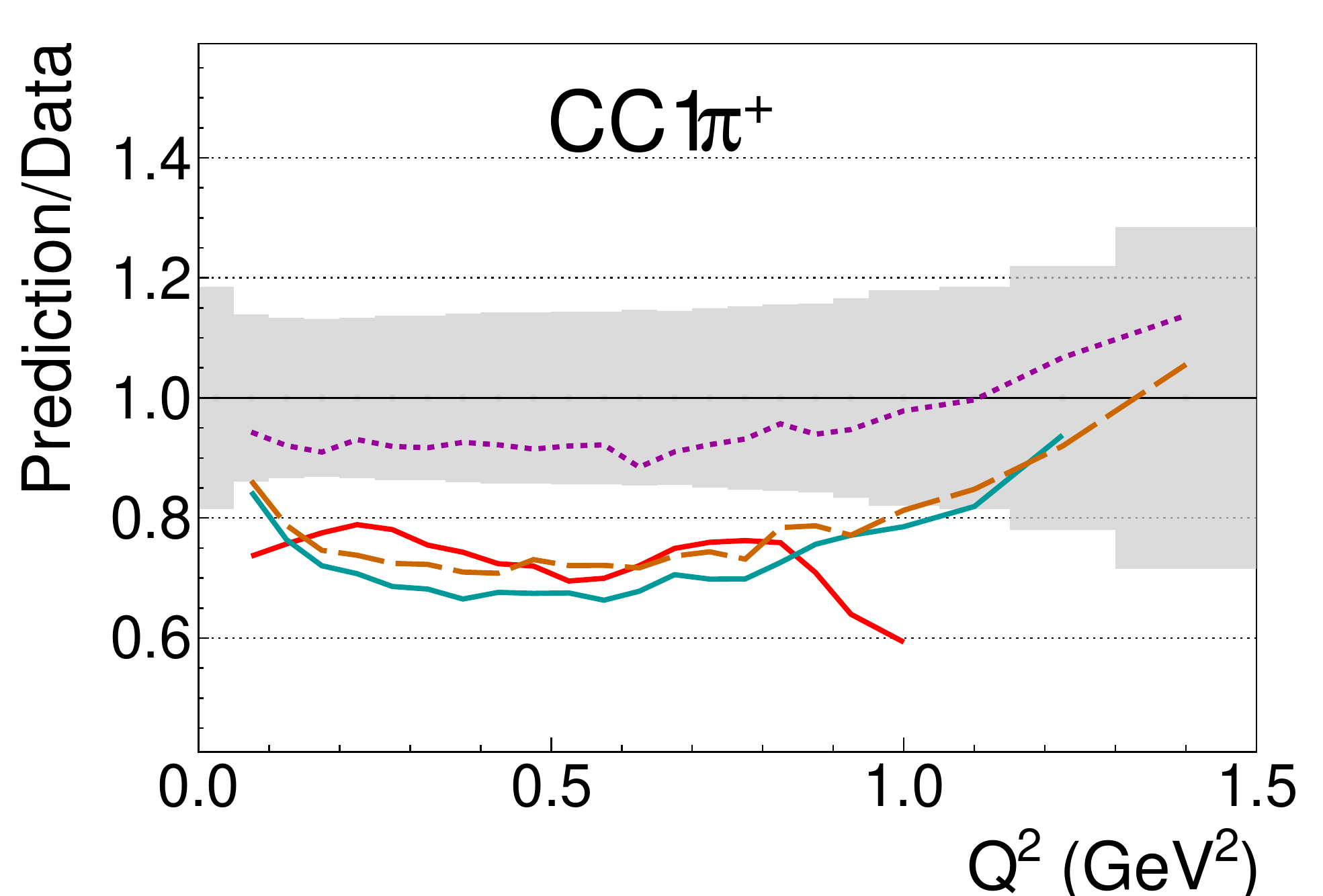}
  \includegraphics[width=0.33\textwidth]{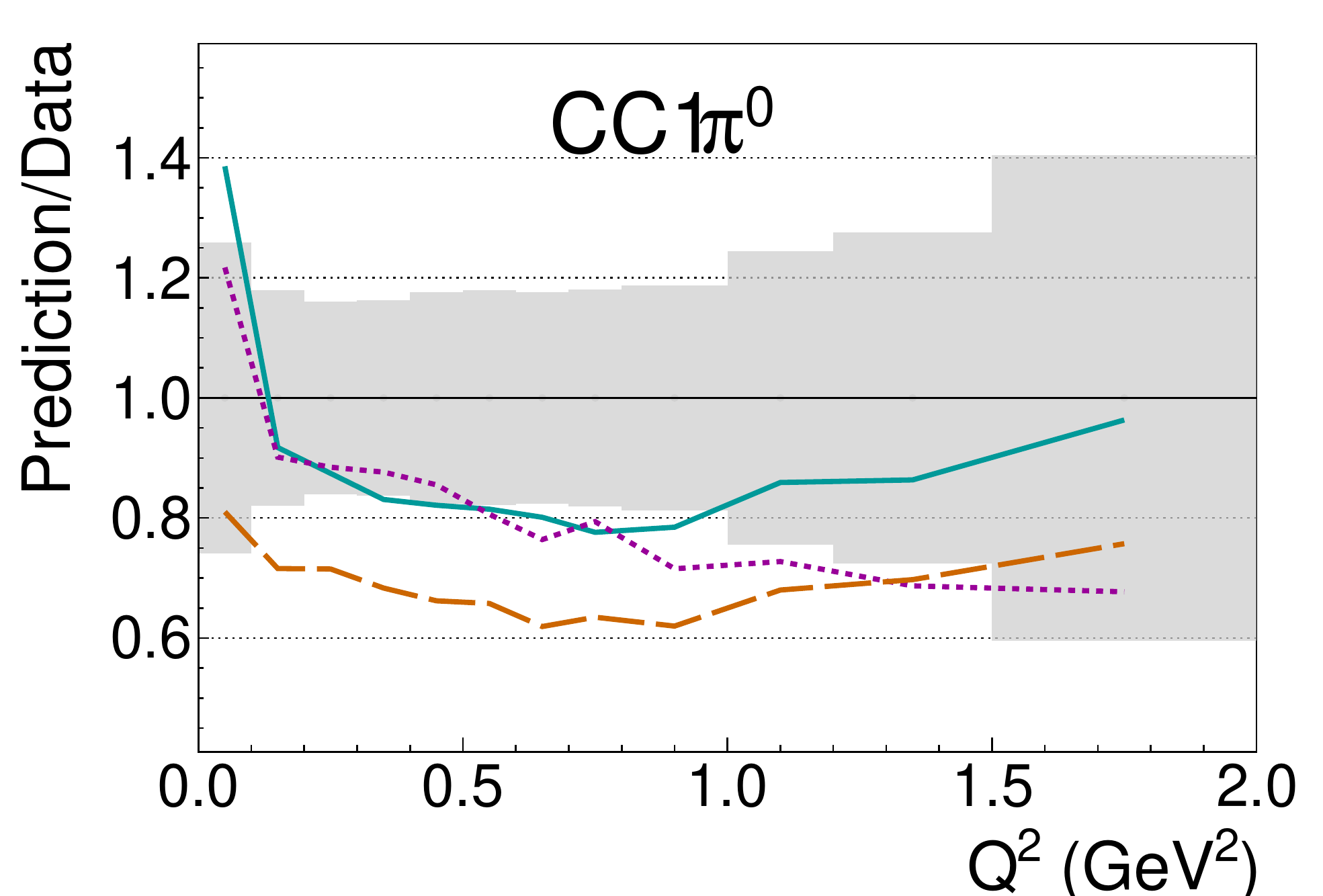}\\
  \includegraphics[width=0.5\textwidth]{horiz_legend}
  \caption{Predictions and \mb data for the flux-averaged differential
    cross section \diff{Q^2} for \ccpip (left) and \ccpi
    (right). Ratios of the predictions to the \mb data are shown in
    the lower row.}
  \label{fig:q2}
\end{ltxfigure}

\begin{ltxfigure}
  \centering
  \includegraphics[width=0.33\textwidth]{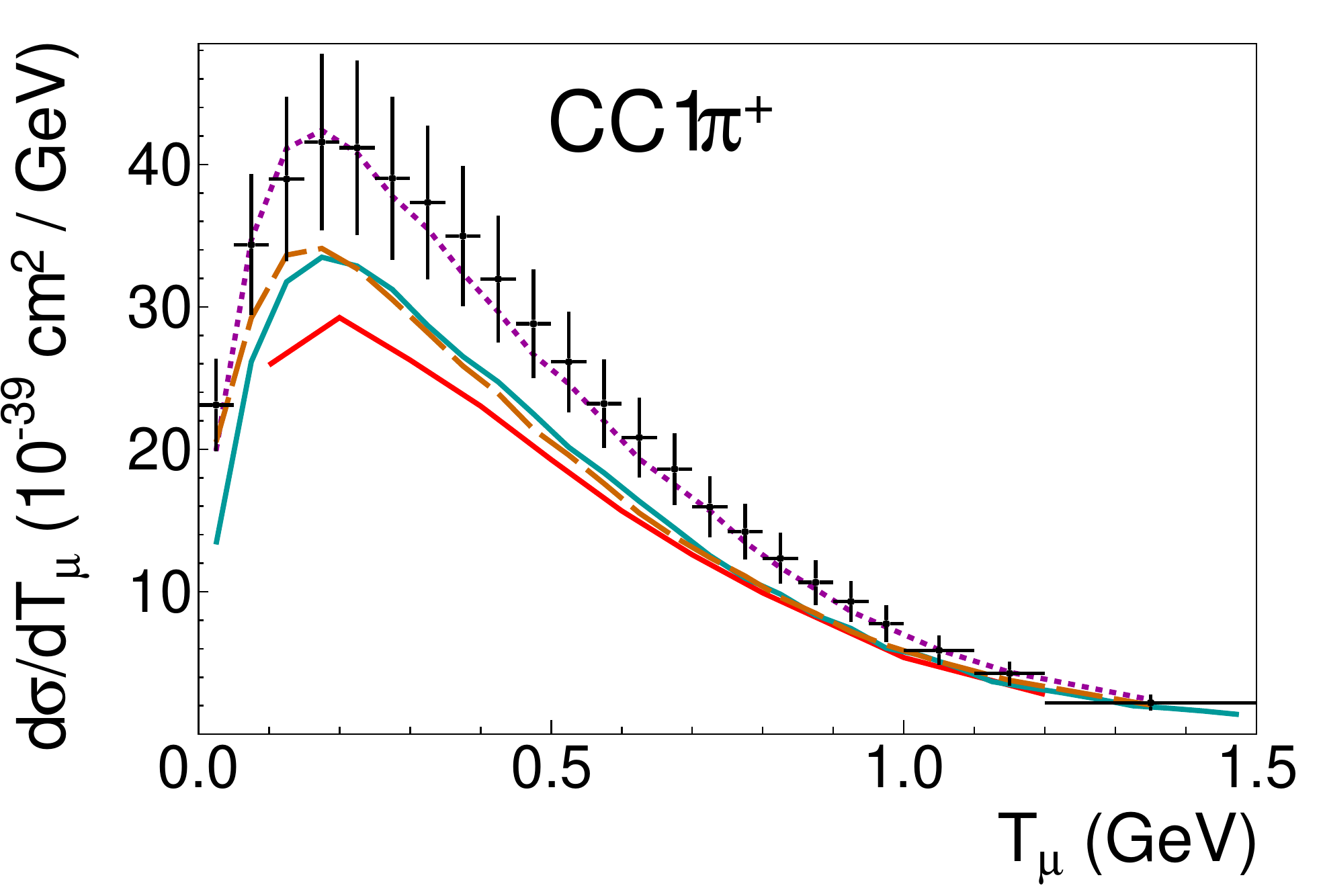}
  \includegraphics[width=0.33\textwidth]{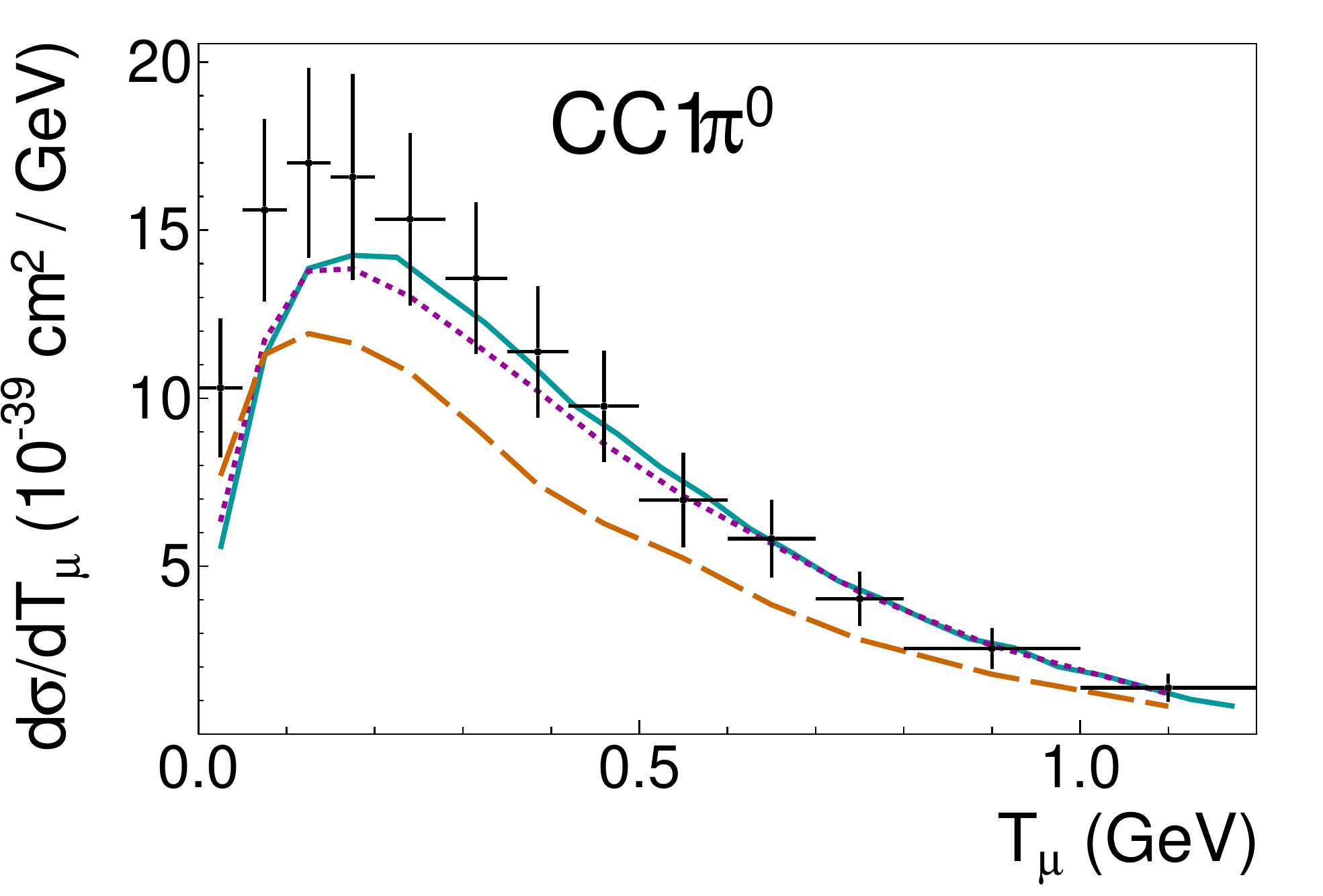}\\
  \includegraphics[width=0.33\textwidth]{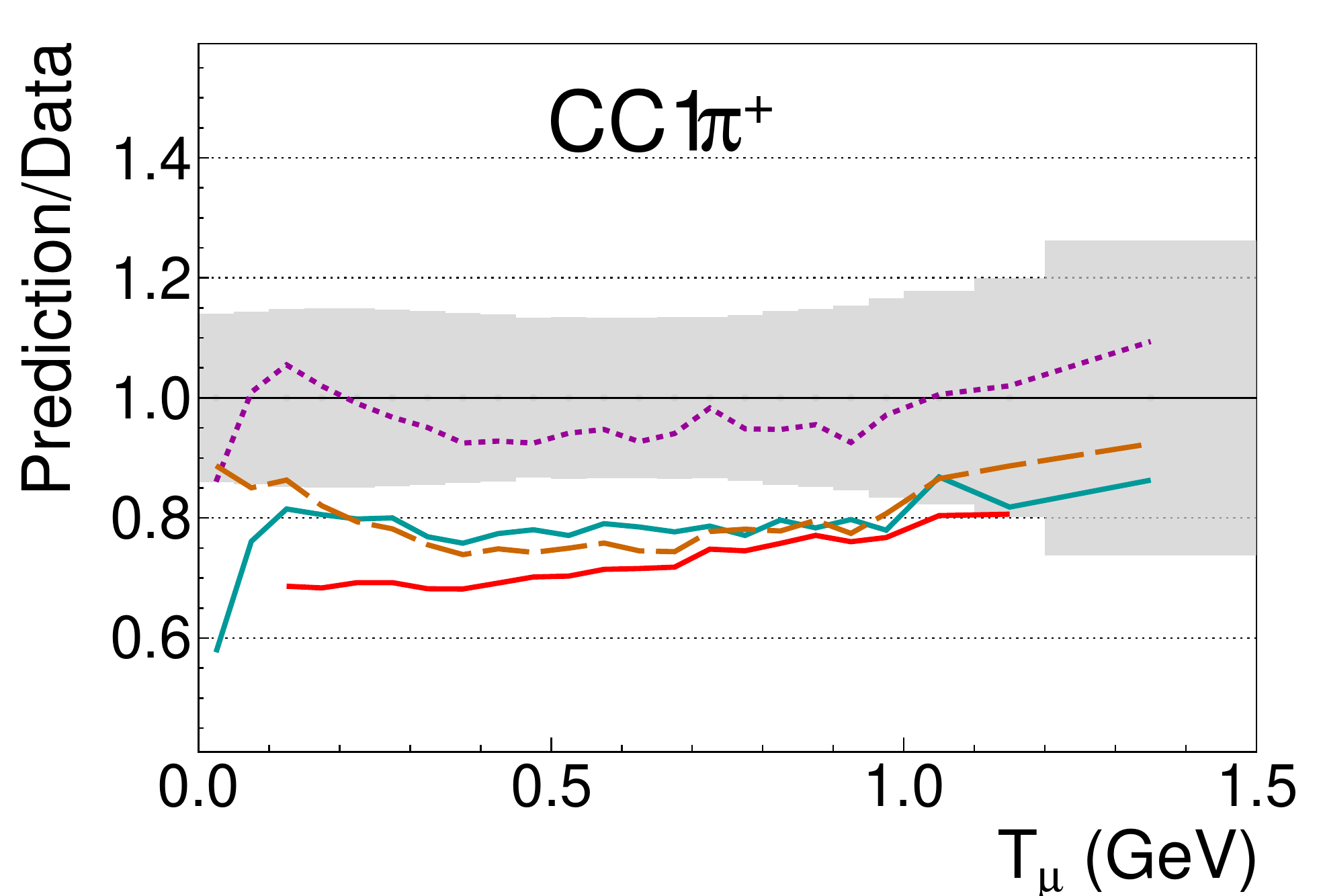}
  \includegraphics[width=0.33\textwidth]{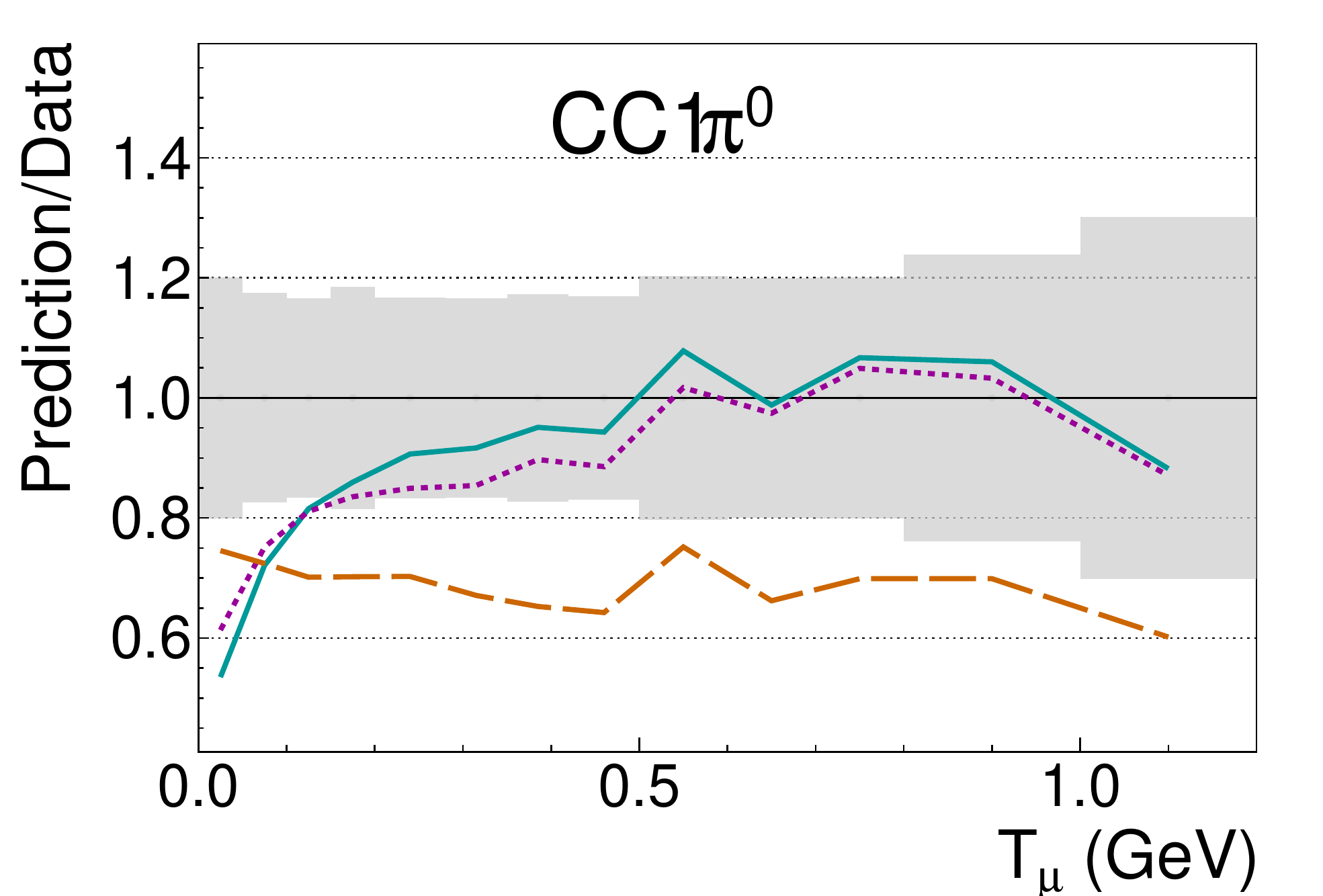}\\
  \includegraphics[width=0.5\textwidth]{horiz_legend}
  \caption{Predictions and \mb data for the flux-averaged differential
    cross section in muon kinetic energy \diff{T_\mu} for \ccpip
    (left) and \ccpi (right). Ratios of the predictions to the \mb
    data are shown in the lower row.}
  \label{fig:tmu}
\end{ltxfigure}

Predictions and \mb data for the differential cross section
\diff{Q^2}, along with ratios, are shown in
Figure~\ref{fig:q2}. Again, the predictions differ in overall
normalization of the cross section by up to 50\%, but show
similarities in shape, with the \mb data having greater strength in
the $Q^2 \approx \unit[0.5]{GeV^2}$ region. The differential cross
section in muon kinetic energy, shown in Figure~\ref{fig:tmu}, is
strongly correlated with $Q^2$, and so shows some differences in the
level of agreement, as compared with the \diff{Q^2} differential cross
section. This suggests some further disagreement in the muon angle
distribution.

\begin{ltxfigure}
  \centering
    \includegraphics[width=0.3\textwidth]{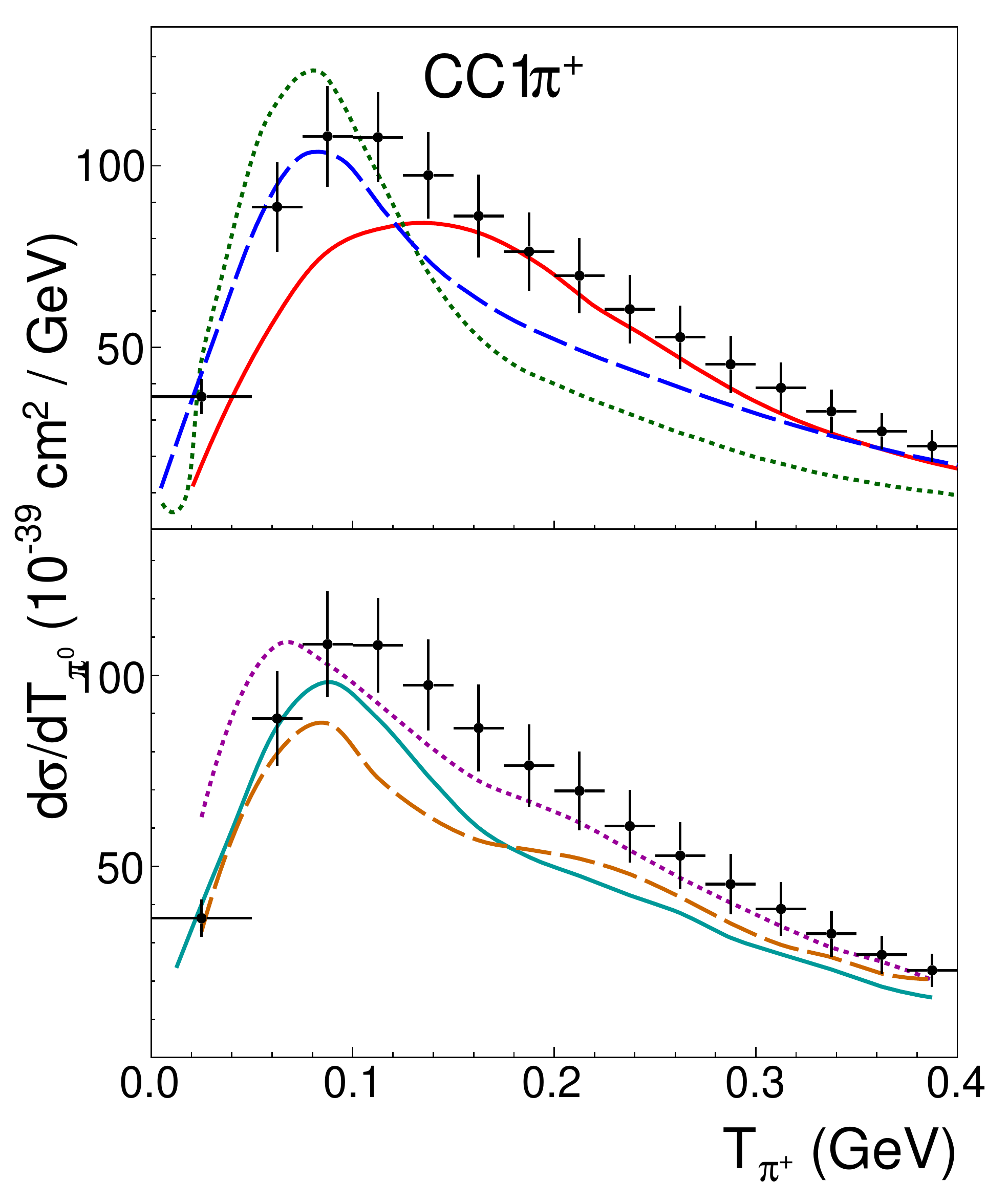}
    \includegraphics[width=0.3\textwidth]{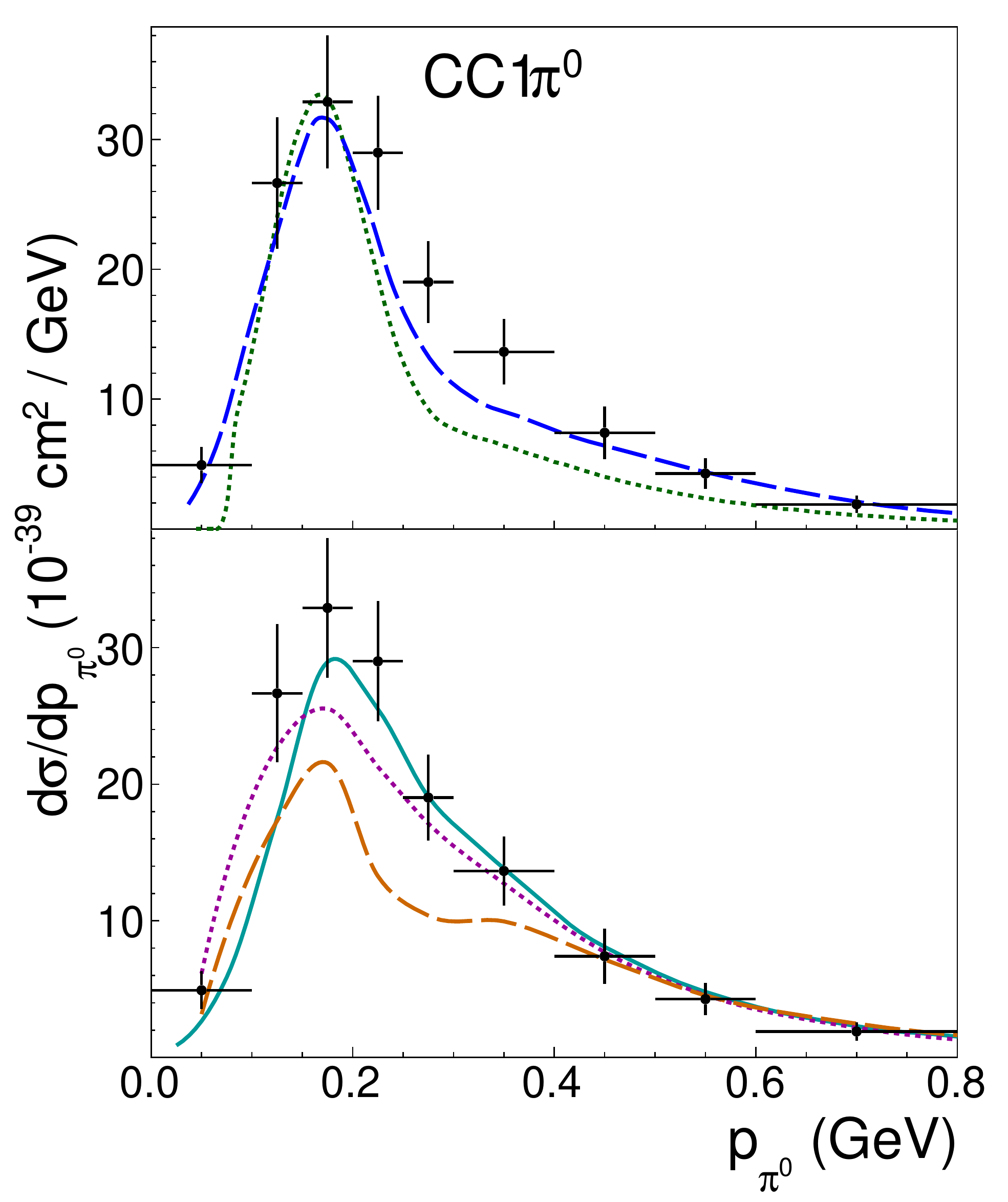}
    \includegraphics[width=0.3\textwidth]{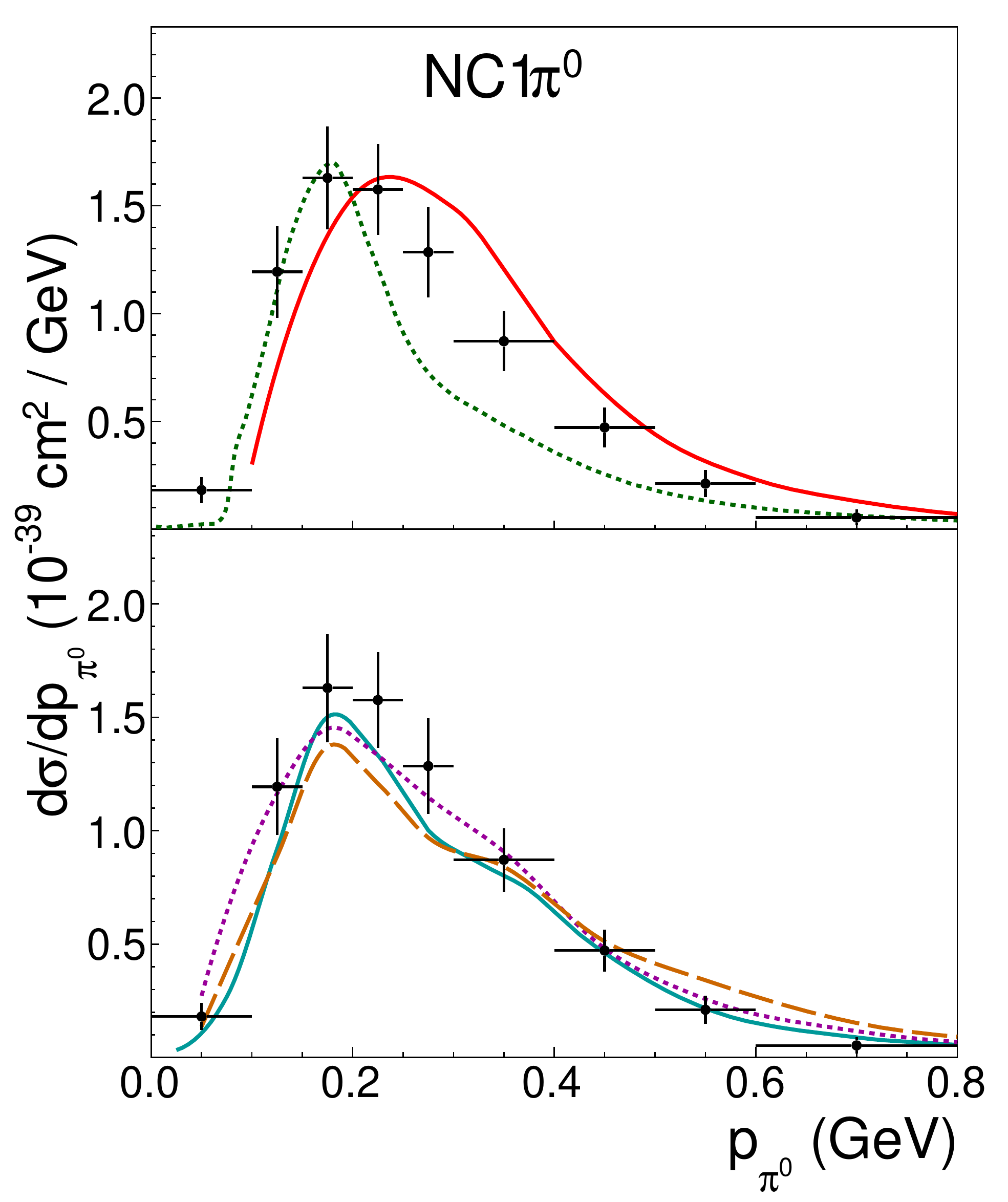}\\
    \includegraphics[width=0.5\textwidth]{horiz_legend}
    \caption{Predictions and \mb data for the differential cross
      sections in pion kinetic energy for \ccpip (left), pion momentum
      for \ccpi (middle), and pion momentum for \ncpi (right).  The
      top row shows the theoretical model predictions, while the
      bottom row shows the MC generator predictions. (The data is the
      same top and bottom.) }
  \label{fig:tpi}
\end{ltxfigure}

Figure~\ref{fig:tpi} shows the differential cross sections in pion
momentum (kinetic energy for \ccpip). These distributions are expected
to be significantly altered by pion reinteractions within the nucleus,
and so potentially provide an important test of final state effects.

There are significant differences in both shape and normalization
between the predictions, and a corresponding difference in level of
agreement with the data. \citet{gibuu-vs-mb} have
drawn attention to the region of pion kinetic energy around
\unit[0.2]{GeV} (pion momentum $p_\pi \approx \unit[0.3]{GeV}$), where
final state effects should reduce the differential cross section
through $\Delta$ production. Most of the models show a dip in this
region for \ccpip, which does not appear to be present in the data.

\begin{ltxfigure}
  \centering
  \includegraphics[width=0.33\textwidth]{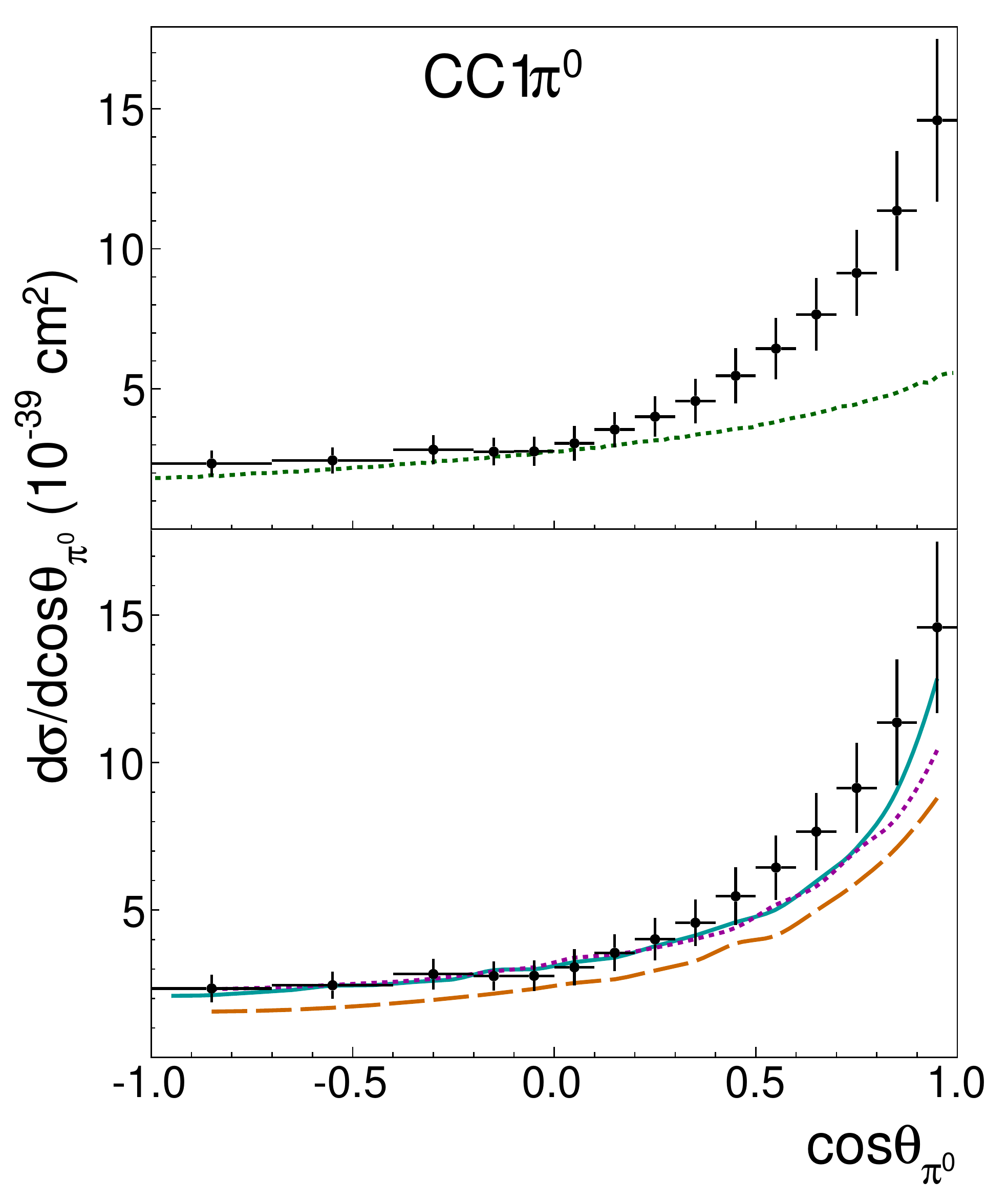}
  \includegraphics[width=0.33\textwidth]{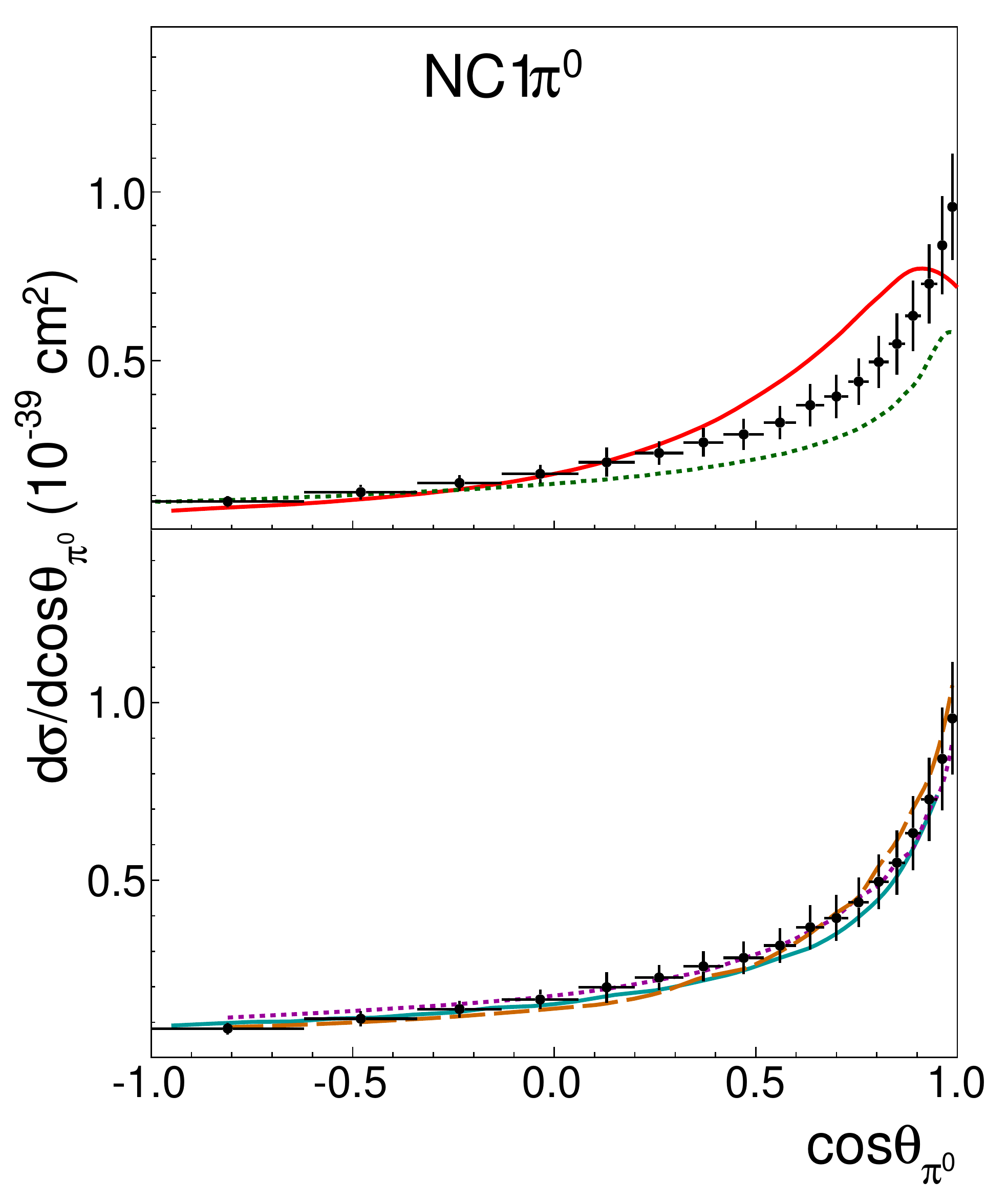}\\
  \includegraphics[width=0.5\textwidth]{horiz_legend}
  \caption{Predictions and \mb data for the differential cross
    section in pion angle \diff{\cos \theta_\pi} for \ccpi (left)
    and \ncpi (right). The top row shows the theoretical model predictions, while
    the bottom row shows the MC generator predictions. (The data is
    the same top and bottom.)}
  \label{fig:cospi}
\end{ltxfigure}

Differential cross sections in pion angle for \ccpi and \ncpi are
shown in Figure~\ref{fig:cospi} (pion angle distributions in \ccpip
are only provided by \mb in the form of double-differential cross
sections over a limited range in $\cos \theta_\pi$ and are not shown
here). While the agreement with data in \ncpi is quite good,
especially for the generators, the cross section is more
forward-peaked in the \ccpi data than for all of the predictions.

\section{Conclusions}

Predictions for neutrino-induced pion production from $\mathrm{CH_2}$
have been compared with the data from \mb. Most models fall below the 
data in total CC cross section, and there are also some shape
disagreements in the differential cross sections. The largest
variation between predictions (and, hence, in the level of agreement
with data) is in the pion momentum distributions, which suggests
differences in final state effects.

The next step in understanding the differences between the predictions
will be to compare predictions for neutrino-induced pion production on
free nucleons, which are free of nuclear and final state effects.

\section{Acknowledgements}

Predictions for the models, along with help in understanding the predictions,
were kindly provided by M.~Athar, S.~Chauhan, S.~Dytman, H.~Gallagher,
T.~Golan, Y.~Hayato, E.~Hernandez, O.~Lalakulich, U.~Mosel, J.~Nieves and
J.~Sobczyk.

\bibliographystyle{aipproc}
\bibliography{nuint12-proceedings}{}

\end{document}